\documentclass[a4paper]{jpconf}
\usepackage{microtype}
\usepackage{supertabular}
\usepackage{array}
\newcolumntype{C}[1]{>{\centering\arraybackslash}m{#1}}
\newcolumntype{R}[1]{>{\raggedleft\arraybackslash}m{#1}}
\usepackage{amsmath}
\usepackage{amsfonts}
\usepackage{amssymb}
\usepackage{amscd}
\usepackage{cancel}
\usepackage[utf8]{inputenc}
\usepackage{ntheorem}
\usepackage[american]{babel}
\usepackage[autostyle]{csquotes}
\usepackage[T1]{fontenc}
\usepackage{textcomp}
\usepackage[dvips]{rotating}
\usepackage{fancybox}
\usepackage{eurosym}
\usepackage{epsfig}
\usepackage[pdftex]{color}
\usepackage{graphicx}
\usepackage{lmodern}
\usepackage{array}
\usepackage{hyperref}

\newcommand{\avg}[1]{\left< #1 \right>} % for average
\renewcommand{\d}[2]{\frac{d #1}{d #2}} % for derivatives
\newcommand{\ket}[1]{\left| #1 \right>} % for Dirac bras
\newcommand{\bra}[1]{\left< #1 \right|} % for Dirac kets
\let\baraccent=\= % rename builtin command \= to \baraccent
\renewcommand{\=}[1]{\stackrel{#1}{=}} % for putting numbers above =

\newcommand{\bmat}{\begin{pmatrix}}
\newcommand{\emat}{\end{pmatrix}}
\newcommand{\bdet}{\begin{dmatrix}}
\newcommand{\edet}{\end{dmatrix}}

\newcommand{\lra}{\longrightarrow}
\newcommand{\Lra}{\Leftrightarrow}

\newcommand{\tn}[1]{\textnormal{#1}}

\newcommand{\refeqn}[1]{Eq.~(\ref{#1})}
\newcommand{\refeqns}[2]{Eqs.~(\ref{#1}) and (\ref{#2})}
\newcommand{\reffig}[1]{Fig.~\ref{#1}}
\newcommand{\reffigs}[2]{Figs.~\ref{#1} and \ref{#2}}
\newcommand{\refsec}[1]{Sec.~\ref{#1}}

\renewcommand{\(}{\left(}
\renewcommand{\)}{\right)}

\newcommand{\Comm}[1]{\Bigl[\,#1\,\Bigr]}

\newcommand{\Commp}[1]{\Bigl[\,#1\,\Bigr]_\mp}
\renewcommand{\i}{\mathrm{i}}
\renewcommand{\d}{\mathrm{d}}

\newcommand{\dint}[2]{\mathrm{d}^{#1}{#2}}
\begin{document}
\title{Few-particle quantum dynamics---comparing Nonequilibrium Green's functions with the generalized Kadanoff-Baym ansatz to density operator theory}

\author{S~Hermanns$^1$, K~Balzer$^2$ and M~Bonitz$^1$}
\address{$^1$ Institut f\"ur Theoretische Physik und Astrophysik,
Universit\"at Kiel, D-24098 Kiel, Germany}
\address{$^2$ University of Hamburg, Max Planck Research Department for Structural Dynamics
Building 99 (CFEL), Luruper Chaussee 149, D-22761 Hamburg, Germany}
\ead{hermanns@theo-physik.uni-kiel.de}
%---
%---
\begin{abstract}
The nonequilibrium description of quantum systems requires, for more than two or three particles, the use of a reduced description to be numerically tractable.
Two possible approaches are based on either reduced density matrices or nonequilibrium Green's functions (NEGF). Both concepts are formulated in terms of hierarchies of coupled equations---the Bogoliubov-Born-Green-Kirkwood-Yvon (BBGKY) hierarchy for the reduced density operators and the Martin-Schwinger-hierarchy (MS) for the Green's functions, respectively. In both cases, similar approximations are introduced to decouple the hierarchy, yet still many questions regarding the correspondence of both approaches remain open.

Here we analyze this correspondence by studying the generalized Kadanoff-Baym ansatz (GKBA) that reduces the NEGF to a single-time theory. Starting from the BBGKY-hierarchy we present the approximations that are necessary to recover the GKBA result both, with Hartree-Fock propagators (HF-GKBA) and propagators in second Born approximation. To test the quality of the HF-GKBA, we study the dynamics of a 4-electron Hubbard nanocluster starting from a strong nonequilibrium initial state and compare to exact results and the Wang-Cassing approximation to the BBGKY hierarchy presented recently by \textit{Akbari et~al.} \cite{akbari12}.
\end{abstract}
%---
%---
\section{Introduction}
%---
%---
The \textit{ab initio} time-dependent description of quantum many-body systems has been a major focus in the physics and chemistry communities since the invention of quantum mechanics. Despite many efforts, up to the present time, the exact (analytic or numerical) solution of the underlying equation of motion for the wavefunction $\ket{\Psi}_\tn{S}(t)$ of the system---the Schrödinger equation---is, in general, restricted to only very few particles, due to the exponentially growing complexity with the particle number, e.g. \cite{bauch08}. To overcome this limitation a great variety of methodologies have been developed to approximately describe systems of larger particle numbers. This includes time-dependent density functional theory \cite{runge84}, time-dependent Hartree-Fock \cite{kulander87,bunge92,doyle68, thouless62,bonche76}, multi-configuration time-dependent Hartree-Fock \cite{hochstuhl10}, time-dependent coupled cluster theory \cite{bartlett81}, M\o{}ller-Plesset many-body perturbation theory \cite{moller34}, density matrix renormalization group based approaches \cite{schollwoeck05} or dynamical mean field theory, e.g. \cite{Georges96}. 

In this article, we focus on two different closely related powerful methods: non-equilibrium Green's functions (NEGF) \cite{kb-book} and reduced density operators (DO), e.g. \cite{born46,kirkwood47,bonitz98,cassing92}, respectively. Both methods involve the solution of a coupled hierarchy---the MS and BBGKY hierarchy, respectively---of reduced quantities, where the solution of the full hierarchy is equivalent to the solution of the $N$-particle Schrödinger equation or von Neumann equation, in the case of pure or mixed states, respectively. To reduce the complexity, both methods aim at simplifications via a physically motivated truncation of the hierarchy. Since both approaches are selfcontained and independent, comparisons of the two are of great interest for applications. NEGF are commonly regarded as more accurate, however, nonequilibrium solutions of the two-time equations of motion---the Keldysh-Kadanoff-Baym equations (KBE)---are computationally very expensive. While there has been remarkable progress during the last decade for homogeneous \cite{kwong00,lorke06} and inhomogeneous systems, e.g. \cite{stan09,balzer10,balzer10a}, the two-time structure puts strong limits on the achievable propagation time. In contrast, the single-time density operator approach does not suffer this problem. 
Furthermore, recent solutions of the KBE for finite systems \cite{puigvonfriesen09,puigvonfriesen10} indicated unphysical long-time behavior (damping). Here, again solutions using single-time equations \cite{hermanns12} are, apparantly, closer to the exact result.
It is, therefore, important to understand how the two-time and single-time approximations to the hierarchy are related to each other and whether and when single-time solutions are justified.

The recipe how to derive the single-time approximation from the NEGF for an arbitrary selfenergy is the so-called generalized Kadanoff-Baym ansatz (GKBA) derived by Lipavsky, Spicka and Velicky \cite{lipavsky86,spicka05}. The GKBA has, so far been used for spatially homogeneous systems, for electron-phonon scattering and for Coulomb scattering in the second Born approximation. Numerical comparisons for the case of Coulomb scattering have shown satisfactory agreement \cite{bonitz96jpcm,binder97,kwong98}. Similar observations have been recently made for finite systems \cite{hermanns12}. Nevertheless, a systematic analysis is still missing. The aim of this paper is two-fold. First, we discuss the relation between NEGF and DO (on the level of the second Born approximation). We take the opposite route, compared to Lipavsky et al. \cite{lipavsky86,spicka05} and derive the GKBA result, starting from the BBGKY-hierarchy and identifying the approximations necessary to obtain the GKBA. Second, we consider, as an example of 
the quantum dynamics of a finite system,  the evolution of a 4-site Hubbard nano-cluster---computed with Green's functions and the GKBA---and compare them to BBGKY-based results by \textit{Akbari~et~al.} \cite{akbari12} using the Wang/Cassing approximation. 
%---
%---
\section{Theory}
%---
%---
In this section, we give a brief overview of the theoretical foundations of the two aforementioned methods for the statistical description of quantum many-body systems. One of the most basic differences between these two approaches is the incorporation of the spin statistics of the particles, which, e.g., for Fermions leads to the effect of Pauli blocking \cite{vetter71}. While the Green's function method, which will be described in \refsec{sec:NEGF}, has the spin statistics intrinsically build in by use of bosonic/fermionic creation- and annhilation-operators, the reduced density operator theory is formulated for spinless particles, which requires an explicit (anti-)symmetrization of the equations. This advantage of the former approach is at least partly counterfeited by the more complicated structure of the equations for the Green's function $G(t,t')$, which depends---apart from the physical time $t$---on an additional time argument $t'$, e.g., providing access to the corresponding ionized system. Numerically, this leads to comparatively more involved calculations with the Green's function approach. To make up for that, the introduction of the GKBA for the single-particle Green's function allows for a simple approximate reconstruction of the two-time Green's function from its time-diagonal value, by which a scaling of computation time comparable to density operator theory is achieved. 
%--- 
\subsection{Density operator theory}
\label{sec:DOT}
%---
For an $N$-particle quantum system in a mixed state (in contact with a stationary environment), the proper fundamental quantity is the $N$-particle density operator
\begin{equation}
\rho_{1\ldots N}=\sum_{k}W_k\ket{\Psi_{1\ldots N}^{(k)}}\bra{\Psi_{1\ldots N}^{(k)}}\,\qquad \sum_k W_k = 1 ,
\end{equation}
where $\ket{\Psi_{1\ldots N}^{(k)}}$ is a possible micro-state compatible with a given macro-state of the whole system including the bath, and $W_k$ denotes the probability of its realization. This description is valid if the interaction of the system with the bath is weak. The density operator obeys the von Neumann equation
\begin{equation}
\i\hbar\partial_t\rho_{1\ldots N}-\Comm{H_{1\ldots N}\,,\,\rho_{1\ldots N}}(t)=0\,,
\label{eq:neumann}
\end{equation}
where $H_{1\ldots N}$ is the Hamiltonian of the $N$-particle system and $\Comm{\cdot\,,\,\cdot\cdot}$ denotes the standard commutator. 
For a given initial state $\rho_{1\ldots N}(t_0)=\rho_{1\ldots N}^{(0)}$, the solution of this equation completely determines the time evolution of the system and  is equivalent to the solution of the Schrödinger equation for all possible micro-states.  
\subsubsection{Nonequilibrium quantum BBGKY-Hierarchy.}
Instead of the full $N$-particle density operator, it is usually useful to consider reduced density operators, the $s$-particle operator $F_{1\ldots s}$ being defined as a partial trace over the $N$-body density operator $\rho_{1\ldots N}$ \cite{bonitz98},
\begin{equation}
F_{1\ldots s}=\mathcal{V}^s\tn{Tr}_{s+1\ldots N}\rho_{1\ldots N}\,,\quad \frac{1}{\mathcal{V}^s}\tn{Tr}_{1\ldots s}F_{1\ldots s}=1\,,
\end{equation}
where $\mathcal{V}$ denotes the volume, and the partial trace $\tn{Tr}_{1\ldots s}A$ of an operator $A$ in an arbitrary basis of states 
$|x\rangle = |x_1\rangle |x_2\rangle \dots |x_N\rangle$ is defined as
\begin{equation}
\tn{Tr}_{1\ldots s}A_{1\dots N}=\sum_{x_1\ldots x_s}A(x_1\ldots x_s,x_{s+1}\ldots x_{N};x_1\ldots x_s,x'_{s+1}\ldots x'_{N})\,.
\end{equation}
The density operators obey a system of equations of motion---the BBGKY hierarchy that follows from taking the partial trace over Eq.~(\ref{eq:neumann})
\begin{eqnarray}
\label{eq:BBGKY1}
\i\hbar \partial_t\, F_1-\Comm{H_1\,,F_1} &=& n\tn{Tr}_2\Comm{V_{12}\,,F_{12}}\,, \\
\i\hbar \partial_t\, F_{12}-\Comm{H_{12}\,,F_{12}}&=& n\tn{Tr}_3\Comm{V_{13}+V_{23}\,,F_{123}}\,,\label{eq:BBGKY2}\\
\dots & \dots & \dots \nonumber\\
F_1(t_0)=F_1^0\,,&&F_{12}(t_0)=F_{12}^0\,, \qquad \dots\nonumber
\end{eqnarray}
and so on. It is obvious that the whole hierarchy is equivalent to the von Neumann equation. Note that this system of equations is local in time, all functions depend on a single physical time $t$, and it has to be complemented by initial conditions $F_1(t_0),F_{12}(t_0)$ and so on. Here $H_1$ is the single-particle Hamiltonian whereas the two-particle Hamiltonian is given by $H_{12}=H_1+H_2+V_{12}$. As mentioned before, \refeqns{eq:BBGKY1}{eq:BBGKY2} are written for spinless particles to make the mathematical structure more transparent. We will add the appropriate exchange contributions in the next section where we consider the (anti-)symmetrization and approximations to the BBGKY hierarchy via a cluster expansion.
\subsubsection{(Anti-)Symmetrization and Cluster expansion of the BBGKY hierarchy.}
For the practical analytical or numerical solution of the BBGKY hierarchy, it is obvious that it has to be truncated to become tractable. A suitable approach consists in performing the so-called cluster expansion to separate the two-particle, three-particle and higher correlations from the ideal part of the  density operator which is given by a product of single-particle operators,
\begin{equation}
\label{eq:F12F123}
\begin{split}
F_{12}(t)&=F_1(t)F_2(t)+c_{12}(t)\,,\\
F_{123}(t)&=F_1(t)F_2(t)F_3(t)  + F_1(t)c_{23}(t) + F_2(t)c_{13}(t)+ F_3(t)c_{12}(t)+c_{123}(t)\,,\\
\end{split}
\end{equation}
and analogously for the higher-order density operators.

To correctly account for the spin statistics of bosons (fermions) we now (anti-)symmetrize all expressions. This can be done by introducing matrix representations with respect to an (anti-)symmetrized system of basis states defined in Fock space. Alternatively, (anti-)symmetric expectation values of observables can be computed with standard Hilbert space states when the respective operators are (anti-)symmetrized \cite{boercker}. Here we follow the latter idea as it leads to more compact expressions. The (anti-)symmetrization of the density operators is achieved by replacing
\begin{equation}
\begin{split}
F_{12}&\lra F_{12}\Lambda^{\pm}_{12}\,,\\
c_{12}&\lra c_{12}\Lambda^{\pm}_{12}\,,\\
F_{123}&\lra F_{12}\Lambda^{\pm}_{123}\,,\\
c_{123}&\lra c_{123}\Lambda^{\pm}_{123}\,,
\end{split}
\end{equation}
where the binary/ternary (anti-)symmetrization operator is defined by its action on an arbitrary two-particle/three-particle state $\ket{12}$ and $\ket{123}$, respectively,
\begin{equation}
\begin{split}
\Lambda^{\pm}_{12}\ket{12}&=(1\pm P_{12})\ket{12}=\ket{12}\pm\ket{21}\,,\\
\Lambda^{\pm}_{123}\ket{123}&=\Lambda^{\pm}_{12}(1\pm P_{13}\pm P_{23})\ket{123}
\end{split}
\end{equation}
and the upper/lower sign applies to bosons/fermions.

We now introduce these cluster expansions (\ref{eq:F12F123}) with the (anti-)symmetrized density operators into the BBGKY-hierarchy (\ref{eq:BBGKY1}\,,\,\ref{eq:BBGKY2}). We limit ourselves to the first three equations where the decoupling is achieved by neglecting 
four-particle correlations, $c_{1234}=0$, (for details, see Ref. \cite{bonitz98}),
\begin{eqnarray}
\label{eq:BBGKY1cor}
\i\hbar \partial_t\, F_1-\Comm{\bar{H}_1^0\,,F_1} &=& n\tn{Tr}_2\Comm{V_{12}\,,\,c_{12}}\Lambda_{12}^\pm\,, \\
\i\hbar \partial_t\, c_{12}-\Comm{\bar{H}_{12}^0\,,c_{12}}&=&\hat{V}_{12}F_1F_2-F_1F_2\hat{V}^\dagger_{12}+n\tn{Tr}_3\Comm{V_{13}+V_{23}\,,\,c_{123}}P_{13;23}\label{eq:BBGKY2cor}  \\
&&\quad +L_{12}+\Pi_{12}\,, \\
\i\hbar \partial_t\, c_{123}-\Comm{\bar{H}_{123}^0\,,c_{123}}&=&\hat{V}^\dagger_{12}F_1F_2F_3+\left(\hat{V}^\dagger_{13}+\hat{V}^\dagger_{23}\right)F_3c_{12}\nonumber\\
&&\quad \mp nF_3\left(F_1V_{13}+F_2V_{23}\right)c_{12}\mp n\left(c_{13}V_{13}+c_{23}V_{23}\right)c_{12}\label{eq:BBGKY3cor}\\
&&\quad +\Pi_{123}+L_{123}+\mathcal{P}_{123}(\tn{rhs.})-\tn{h.c.}(\tn{rhs.})\nonumber \,,\\
%c_{1234} &\equiv & 0\,,\\
F_1(t_0)&=&F_1^0\,,\quad c_{12}(t_0)=c_{12}^0\,,\quad c_{123}(t_0)=c_{123}^0, \qquad P_{13;23}=(1\pm P_{13}\pm P_{23})\,,\nonumber\\
L_{12},L_{123} &=& \tn{ladder terms}\,,\qquad \Pi_{12},\Pi_{123} = \tn{polarization terms}\,,\nonumber\\
\mathcal{P}_{123}(\tn{rhs.}) &=& \tn{cyclic permutation of 1, 2, 3 in all terms on the rhs.}\,, \nonumber\\
\tn{h.c.}(\tn{rhs.}) &=& \tn{hermitean conjugate of all terms on rhs. including } \mathcal{P}_{123}(\tn{rhs.}) \nonumber\,.
\end{eqnarray}
Since below we will focus on the second Born approximation we do not explicitly write out the ladder and polarization terms \cite{bonitz98} since they will be neglected. On the left-hand sides we introduced mean field Hamiltonians which are renormalized by a Hartree-Fock potential $U^\tn{HF}$,
\begin{eqnarray}
\bar{H}_1^0 &=& H_1+U_1^\tn{HF}\,,\qquad U_1^\tn{HF} = n\tn{Tr}_2V_{12}F_2\Lambda_{12}^\pm\,,\\
\bar{H}_{12}^0 &=& \bar{H}_1^0+\bar{H}_2^0\,,\quad\quad \;\,\bar{H}_{123}^0 = \bar{H}_1^0+\bar{H}_2^0+\bar{H}_3^0
\label{eq:H120}
\end{eqnarray}
and the non-hermitian operator $\hat{V}_{12} = (1\pm nF_1\pm nF_2)V_{12}$ which takes into account the exchange renormalization of the pair interaction giving rise, e.g., to Pauli blocking.

Below we will consider two approximations: first, $c_{123}\equiv 0$, which leads to the second Born (2B) approximation of NEGF theory together with the Hartree-Fock GKBA (HF-GKBA, i.e. with HF propagators), see Sec.~\ref{sec:NEGF}. Second, we include in $c_{123}$ all relevant terms that give rise to 2B plus GKBA with full propagators (HF plus 2B). We start with the second approximation since the first follows from it as a special case. To this end, we solve the third hierarchy equation by retaining on the r.h.s. only terms that are proportional to $c_{12}$:
\begin{equation}
\label{eq:Selfenergyterms}
\left\{\left(\hat{V}^\dagger_{13}+\hat{V}^\dagger_{23}\right)F_3\mp nF_3\left(F_1V_{13}+F_2V_{23}\right)\mp n\left(c_{13}V_{13}+c_{23}V_{23}\right)\right\}c_{12}
-\tn{h.c.}
\end{equation} 
These terms describe the coupling of the pair  1--2 to third particles including medium effects which will give rise to energ renormalization (selfenergy). This expression can be rewritten using the definition of $\hat V$, and neglecting the ladder-type corrections involving products $c_{13}V_{13}$ and $c_{23}V_{23}$, yielding
\begin{equation}
\begin{split}
&\left\{\left(1\pm nF_1\right)\left(1\pm nF_3\right)V_{13}F_3\mp F_1F_3V_{13}\left(1\pm nF_3\right)\right\}c_{12}-\tn{h.c.}+(1 \Lra 2)\\
\quad &=: \left(S_{13}^>\mp S_{13}^<+S_{23}^> \mp S_{23}^< \right)c_{12}-\tn{h.c.}\,,
\end{split}
\label{eq:2b_selfenergy}
\end{equation}
where we defined
\begin{equation}
S_{ab}^\gtrless :=F_a^\gtrless F_b^\gtrless V_{ab}F_b^\gtrless\,. 
\end{equation}
We now rewrite Eq.~(\ref{eq:BBGKY2cor}) in static second Born approximation, neglecting all polarization and ladder terms and retaining, on the r.h.s., only the selfenergy contributions \ref{eq:2b_selfenergy},
\begin{equation}
\i\hbar \partial_t\, c_{123}-\left\{\bar{H}_{123}^{0,\tn{eff}}c_{123}-c_{123}\bar{H}_{123}^{0,\tn{eff}\dagger}\right\}=\left(S_{13}^>\mp S_{13}^<+S_{23}^>\mp S_{23}^<\right)c_{12}-\tn{h.c.}\,,
\end{equation}
where
% we additionally defined $G_{ab}:=nc_{ab}V_{ab}$. 
we have introduced an effective (non-hermitian) three-particle Hamiltonian $\bar{H}_{123}^{0,\tn{eff}}=\bar{H}_{1}+\bar{H}_{2}+\bar{H}_{3}$. We can now formally solve this equation for $c_{123}$ in terms of $c_{12}$,
\begin{equation}
\begin{split}
\label{eq:c123t}
c_{123}(t)&=U_{123}^{0+}(tt_0)c_{123}^0U_{123}^{0-}(t_0t)\\
\qquad &+\frac{1}{\i\hbar}\int_{t_0}^t\d \, \bar{t}\,U_{123}^{0+}(t\bar t)\left\{ \left(S_{13}^>\mp S_{13}^<+S_{23}^>\mp S_{23}^<\right)c_{12}-\tn{h.c.}\right\}\bigg|_{\bar{t}}U_{123}^{0-}(\bar t t)\,, 
\end{split}
\end{equation} 
where we introduced the propagators $U_{123}^{0\pm}$ with the properties
\begin{equation}
\label{eq:U1230}
U_{123}^{0\pm}(tt')=\left[U^{0\mp}_{123}(t't)\right]^\dagger, \qquad U_{123}^{0\pm}(tt')=U_1^{\pm}(tt')U_2^{\pm}(tt')U_3^{\pm}(tt')\,.
\end{equation}
The single-particle propagators obey effective one-particle Schrödinger-type equations
\begin{eqnarray}
\label{eq:U10+}
\left\{\i\hbar\partial_t-\bar{H}_1(t)\right\}U_1^{+}(tt')&=&0\,,\qquad U_1^{+}(tt)=0\,,\\
U_1^{-}(t't)\left\{\i\hbar\partial_{t'}+\bar{H}_1^\dagger(t')\right\}&=&0\,,\qquad U_1^{-}(tt)=0\,,
\label{eq:U10-}
\end{eqnarray}
where, in the second equation, the time derivative acts onto operators placed left of it. The definition of the effective one-particle Hamiltonian $\bar{H}_1$ will be derived in the following.

We now turn to the second hierarchy equation (\ref{eq:BBGKY2cor}), inserting the formal result for $c_{123}(t)$ on the r.h.s. and again neglect the polarization and ladder terms, 
\begin{equation}
\label{eq:c12eom1}
\begin{split}
&\i \hbar \partial_t c_{12}-\Comm{\bar{H}_1^0+\bar{H}_2^0\,,\,c_{12}} = I_{12}^> - I_{12}^<
+ n\tn{Tr}_3\left\{\Comm{U_{123}^{0+}(tt_0)c_{123}^0U_{123}^{0-}(t_0t)}\right\}\Lambda_{123}^\pm \\
\qquad &+\frac{n}{\i\hbar}\int\limits_{t_0}^t\d \bar{t} \,\tn{Tr}_3
\bigg\{
\Comm{V_{13}+V_{23}\,,\,U_{123}^{0+}(t\bar t)\left\{\left(S_{13}^>\mp S_{13}^<+S_{23}^>\mp S_{23}^<\right)c_{12}-\tn{h.c.}\right\}\bigg|_{\bar{t}}U_{123}^{0-}(\bar t t)} \bigg\} \Lambda_{123}^\pm\,.
\end{split}
\end{equation}
Here, the first term on the r.s.h. of \refeqn{eq:BBGKY2cor} has been transformed according to
\begin{equation}
\begin{split}
\hat{V}_{12}F_1F_2-F_1F_2\hat{V}^\dagger_{12}=\left(1\pm nF_1\right)\left(1\pm nF_2\right)V_{12}F_1F_2-F_1F_2V_{12}\left(1\pm nF_1\right)\left(1\pm nF_2\right)=:
I_{12}^> - I_{12}^<\,,
\end{split}
\end{equation}
where we introduced the greater- and less-collision integral operators, $I_{12}^>$ and $I_{12}^<$
\begin{equation}
\begin{split}
I_{ab}^\gtrless &=F_a^\gtrless F_b^\gtrless V_{ab}F_a^\lessgtr F_b^\lessgtr \,,\\
F_a^< &=F_a\,,\\
F_a^> &=1\pm nF_a\,.
\end{split}
\end{equation}   
Inspection of the integral term in \refeqn{eq:c12eom1} reveals that it has the structure of a selfenergy operator, $\tilde{\Sigma}_{12}$, acting on $c_{12}$:
\begin{equation}
\tilde{\Sigma}_{12}(t)c_{12}(t)=\int_{t_0}^t\d \bar{t}\left\{\Sigma^+_{12}(t\bar t)c_{12}(\bar t)U_{12}^-(\bar t t)-\tn{h.c.}\right\}\,,
\end{equation}
which we can decompose into one- and two-particle contributions,
\begin{equation}
\tilde{\Sigma}_{12}=\tilde{\Sigma}_{1}+\tilde{\Sigma}_{2}+\tilde{\Sigma}_{12}^\tn{cor}\,,\qquad \Sigma^\pm_{12}=\Sigma^\pm_{1}U^\pm_{2}+\Sigma^\pm_{2}U^\pm_{1}+\Sigma^{\pm,\tn{corr}}_{12}
\end{equation}
which are, in turn, given by (cf. \refeqn{eq:c12eom1}),
\begin{equation}
\begin{split}
\Sigma_1^+(t \bar t)&=\frac{n}{\i\hbar}\tn{Tr}_3\left\{V_{13}U_{13}^{0+}(t\bar t)\left(S_{13}^>\mp S_{13}^<\right)\bigg|_{\bar{t}}U_3^-(\bar t t)\right\}\Lambda^\pm_{13}\,,\\
\Sigma_2^+(t \bar t)&=\frac{n}{\i\hbar}\tn{Tr}_3\left\{V_{23}U_{23}^{0+}(t\bar t)\left(S_{23}^>\mp S_{23}^<\right)\bigg|_{\bar{t}}U_3^-(\bar t t)\right\}\Lambda^\pm_{23}\,,\\
\Sigma_{12}^{\tn{cor}+}(t \bar t)&=\frac{n}{\i\hbar}\tn{Tr}_3\left\{V_{23}U_{123}^{0+}(t\bar t)\left(S_{13}^>\mp S_{13}^<\right)\bigg|_{\bar{t}}U_3^-(\bar t t)\right\}\Lambda^\pm_{123}+1 \Lra 2\,.
\end{split}
\end{equation}

With this, we can rewrite \refeqn{eq:c12eom1} by collecting all terms acting on $c_{12}$ into an effective two-particle hamiltonian,
\begin{equation}
\label{eq:c12eom2}
\i\hbar \partial_t c_{12}(t)-\left\{H_{12}^{0,\tn{eff}}(t)c_{12}(t) - c_{12}(t)H_{12}^{0,\tn{eff}}(t)\right\}=I_{12}^>(t) - I_{12}^<(t)\,,
\end{equation}
with the definition 
\begin{equation}
H_{12}^{0,\tn{eff}}(t)c_{12}(t) = \bar{H}_{12}^0(t)c_{12}(t)+\int_{t_0}^t\d \bar{t}\,\Sigma^+_{12}(t\bar t)c_{12}(\bar t)U_{12}^{0-}(\bar t t).
\end{equation}
This hamiltonian consists of three parts, $H_{12}^{0,\tn{eff}} = \bar{H}_{1}+\bar{H}_{2}+\bar{H}_{12}^{0\tn{cor}}\,$,
\begin{equation}
\begin{split}
\bar{H}_{1}(t)c_{12}(t)&=\bar{H}_{1}^0(t)c_{12}(t)+\int_{t_0}^t\d \bar{t}\,\Sigma^+_{1}(t\bar t)U^+_{2}(t\bar t)c_{12}(\bar t)U_{12}^{0-}(\bar t t)\,,\\
\bar{H}_{2}(t)c_{12}(t)&=\bar{H}_{2}^0(t)c_{12}(t)+\int_{t_0}^t\d \bar{t}\,U^+_{1}(t\bar t)\Sigma^+_{2}(t\bar t)c_{12}(\bar t)U_{12}^{0-}(\bar t t)\,,\\
\bar{H}_{12}^{0\tn{cor}}(t)c_{12}(t)&=\int_{t_0}^t\d \bar{t}\,\Sigma^{\tn{cor}+}_{12}(t\bar t)c_{12}(\bar t)U_{12}^{0-}(\bar t t)\,.
\end{split}
\end{equation}
To preserve the additivity of $H_{12}^{0,\tn{eff}}$ resulting from the Born approximation, it is necessary to neglect the term $\bar{H}_{12}^{0\tn{cor}}c_{12}$, which yields a result that is consistent with our previous definition of the effective three-particle hamiltonian,
\begin{equation}
\begin{split}
H_{12}^{0,\tn{eff}}c_{12}  &= \(\bar{H}_{1}+\bar{H}_{2}\)c_{12}\,,\\
H_{123}^{0,\tn{eff}}c_{123}&=\(\bar{H}_{1}+\bar{H}_{2}+\bar{H}_{3}\)c_{123}\,.
\end{split}
\end{equation}
This result can now be used to solve the equations of motion, \refeqn{eq:U10+}, for the renormalized one-particle propagator $U_1^+(tt')$ which transforms into
\begin{equation}
\label{eq:DysonU}
\begin{split}
\left\{\i\hbar\partial_t-\bar{H}_{1}^0\right\}U_1^{+}(tt')-\int_{t_0}^t\d \bar{t}\,\Sigma^+_{1}(t\bar t)U_1^{+}(\bar tt')=0\,.
\end{split}
\end{equation}
Using the definition of $S^\gtrless$, we can write out the selfenergy $\Sigma_1^+$ in second Born approximation explicitly (neglecting initial value terms),
\begin{equation}
\begin{split}
\Sigma_1^+(tt')=\frac{n}{i\hbar}\tn{Tr}_3\left\{V_{13}U_1^+(tt')U_3^+(tt')\left[F_1^>F_3^>V_{13}F_3^<\mp F_1^<F_3^<V_{13}F_3^>\right]U_3^-(t't)\bigg|_{t'}\right\}\,.
\end{split}
\end{equation}
The structure of $\Sigma^+_{1}$ suggests to define new quantities
\begin{equation}
\begin{split}
g_a^>(tt')=U_a^+(tt')F_a^>(t')-F_a^>(t)U_a^-(tt')\,,\\
g_a^<(t't)=U_a^+(tt')F_a^<(t')-F_a^<(t)U_a^-(tt')\,,
\end{split}
\label{eq:gkba-do}
\end{equation}
giving rise to a compact and symmetric expresssion
\begin{equation}
\label{eq:Sigma1final}
\begin{split}
\Sigma_1^+(tt')&=\frac{n}{i\hbar}\tn{Tr}_3 V_{13}\bigg\{g_1^>(tt')g_3^>(tt')V_{13}
g_3^<(t't)\mp g_1^<(tt')g_3^<(tt')V_{13}g_3^>(t't)\bigg\}\\
&=:\Sigma_1^>(tt')\mp\Sigma_1^<(tt')\,,
\end{split}
\end{equation}
where we defined the greater- and less-selfenergy $\Sigma_1^>(tt')$ and $\Sigma_1^<(tt')$. With these definitions one can finally write down the equation of motion for the single-particle density operator $F_1$, which reads according to \refeqn{eq:BBGKY1cor},
\begin{equation}
\label{eq:neom}
\begin{split}
&\i\hbar \partial_t\, F_1-\Comm{\bar{H}_1^0\,,F_1} = n\tn{Tr}_2\Comm{V_{12}\,,\,c_{12}}\Lambda_{12}^\pm\\
&\quad=\left(n\tn{Tr}_2\left\{V_{12}U_{12}^{0+}(t,t_0)c_{12}^0U_{12}^{0-}(t_0,t)\right\}-n\tn{Tr}_2\left\{U_{12}^{0+}(t,t_0)c_{12}^0U_{12}^{0-}(t_0,t)V_{12}\right\}\right)\Lambda_{12}^\pm\\
&\quad\quad+\frac{in}{\hbar}\int_{t_0}^t\d \bar{t}\tn{Tr}_2
\bigg(\big\{ V_{12}U_{12}^{0+}(t\bar t)\left[F_1^>F_2^>V_{12}F_1^<F_2^<-F_1^<F_2^<V_{12}F_1^>F_2^>\right]\big|_{\bar{t}}\,U_{12}^{0-}(\bar t t)\Lambda_{12}^\pm\big\}\\
&\qquad\quad\qquad\qquad-\big\{U_{12}^{0+}(t\bar t)\left[F_1^>F_2^>V_{12}F_1^<F_2^<-F_1^<F_2^<V_{12}F_1^>F_2^>\right]\big|_{\bar{t}}\,U_{12}^{0-}(\bar t t) V_{12}\Lambda_{12}^\pm\big\}\bigg) \\
&\quad=\int_{t_0}^t\dint{}{\bar{t}}\,\left\{\Sigma_1^>(t,\bar{t})g^<(\bar{t},t)-\Sigma_1^<(t,\bar{t})g^>(\bar{t},t)\right\}\,.
\end{split}
\end{equation}
Here we have used, after the first line, the formal solution $c_{12}(t)$ of Eq.~(\ref{eq:c12eom2}) that includes initial correlations, $c_{12}^0$ and scattering contributions (the time integral).

In the next section, we will see, that this equation of motion for the reduced single-particle density operator has the same form as the result obtained within the NEGF formalism for the time-diagonal limit of the function $G^<$ in second Born approximation, after applying the GKBA. Thereby it turns out that the full result, as derived above, corresponds to the GKBA with full second-Born propagators. In contrast, the HF-GKBA follows if the propagator equation (\ref{eq:DysonU}) is solved after neglecting the selfenergy correction, which is equivalent to decoupling the BBGKY-hierarchy by requiring $c_{123}=0$, cf. Sec.~\ref{sec:NEGF}.
\subsection{Nonequilibrium Green's functions}
\label{sec:NEGF}
In contrast to density operator theory, presented in the preceding section, the basic quantities in the nonequilibrium Green's function formalism are the creation/annihilation operators $a_i$ and $a^\dagger_i$, which create/annihilate a particle in the $i$-th one-particle orbital and obey the canonical commutator/anticommutator relations for bosonic or fermionic particles,
\begin{equation}
\label{eq:commutatorrelations}
\Commp{\hat c_i^{(\dagger)},\,\hat c_j^{(\dagger)}}=0, \quad \Commp{c_i,\,\hat c_j^\dagger}=\delta_{i,j}\,. 
\end{equation}   
With this, one can define the ensemble average of the combination $N$ creation and annihilation operators at $2N$ different points in time,
\begin{equation}
\begin{split}
&\avg{c_{j_1}(t'_1)\ldots c_{j_N}(t'_N)c_{i_N}^\dagger(t_N)\ldots c_{i_1}^\dagger(t_1)}_\tn{} =: G_{i_1\ldots i_N;j_1\ldots j_N}^{N,<}(t_1\ldots t_N;t'_1\ldots t'_N)\\
&=\tn{Tr }\big\{\rho_\tn{}c_{j_1}(t'_1)\ldots c_{j_N}(t'_N)c_{i_N}^\dagger(t_N)\ldots c_{i_1}^\dagger(t_1)\big\}\,,
\end{split}
\end{equation} 
as the less-part of the real-time $N$-particle thermal Green's function $G^{N,<}$. Here $\rho_\tn{}$ is the density operator of the system and the notation \enquote{$<$} refers to the particular ordering of the operators. For equal times ($t_1=\ldots=t_N=t'_1=\ldots=t'_N$) the quantity $\i^NG^{N,<}$ is just the $N$-particle density operator $\rho_{1\ldots N}$ in the one-particle orbital basis. Analogous to the density operator theory, one can define reduced Green's functions involving fewer operators, so that the most basic quantity, the single-particle two-time less Green's function $G^<_{ij}(t_1,t_2)$ can be defined as
\begin{equation}
G^<_{ij}(t_1,t_2)=\avg{c^\dagger_{j}(t_2)c_{i}(t_1)}_\tn{}\,.
\end{equation}
The Green's functions also obey a hierarchy of equations of motion, the Martin-Schwinger hierarchy (MSH) \cite{martin59}, which for all $N$ connects the $N$-particle Green's function to the $(N-1)$- and the $(N+1)$-particle Green's functions. Since the complexity of the whole hierarchy---due to the additional time-arguments--- is computationally even more demanding, one again resorts to the closure of the hierarchy equation on the single-particle level by a cluster expansion, stating the two-particle Green's function in terms of the one-particle Green's function by introduction of a suitable selfenergy functional $\Sigma(G)$. With this, the equation of motion for the less Green's function attains the well known form of the Keldysh/Kadanoff-Baym equation (KBE),
\begin{equation}
\label{eq:KKBE}
\begin{split}
\i\partial_{t_1}G_{ij}^<(t_1,t_2)&=\sum_kh_{ik}(t_1)G_{kj}^<(t_1,t_2)+\int\mathrm{d}t_3\,\Sigma_{ik}^\tn{R}(t_1,t_3)G_{kj}^<(t_3,t_2)+\int\mathrm{d}t_3\,\Sigma_{ik}^<(t_1,t_3)G_{kj}^\tn{A}(t_3,t_2)\,. \\
\end{split}
\end{equation} 
The quantities $G^R$ and $G^A$---the retarded and advanced Green's functions---are defined as
\begin{align}
G_{ij}^\tn{R/A}(t_1,t_2)&=\pm\theta\(\pm(t_1-t_2)\)\left[G_{ij}^>(t_1,t_2)-G_{ij}^<(t_1,t_2)\right]\,,\\
G^>_{ij}(t_1,t_2)&= -\mathrm{i}\avg{\hat c_i(t_1)\hat c_j^\dagger(t_2)}\,,
\end{align}
where $h_{ij}$ is the $ij$-th matrix element of the Hartree-Fock part of the Hamiltonian. The corresponding components of the selfenergy read in the second order Born approximation:
\begin{equation}
\begin{split}
\Sigma_{ij}^{\tn{2B},<}(t_1,t_2)&=\sum_{klmnrs}w_{ikms}(\sigma\,w_{rnlj}-w_{rlnj})G^<_{kl}(t_1,t_2)G^<_{mn}(t_1,t_2)G^>_{rs}(t_2,t_1)\,,\\
\Sigma_{ij}^\tn{2B,R}(t_1,t_2)&=\sum_{klmnrs}w_{ikms}(\sigma\,w_{rnlj}-w_{rlnj})G^\tn{R}_{kl}(t_1,t_2)G^\tn{R}_{mn}(t_1,t_2)G^\tn{A}_{rs}(t_2,t_1)\,,
\end{split}
\end{equation}
where $\sigma=\{1,2\}$ for spin-polarized/spin-restricted systems. As an alternative formulation of the KBE \textit{Lipavskii et al.} proposed \cite{lipavsky86}
\begin{equation}
\label{eq:gless_lip}
\begin{split}
G^<(t_1,t_2) &= \int_{t_2}^{t_1}\d t_3\,\int_{t_0}^{t_2}\d  t_4\, G^\tn{R}(t_1, t_3)\Sigma^<(t_3,  t_4)G^\tn{A}(t_4,t_2)\\ 
&+\Theta(t_1-t_2)\left[-G^\tn{R}(t_1,t_2)\rho(t_2)+\int_{t_2}^{t_1}\d t_3\,\int_{t_0}^{t_2}\d  t_4\,G^\tn{R}(t_1, t_3)\Sigma^\tn{R}(t_3,  t_4)G^<(t_4,t_2)\right]\\\
&+\Theta(t_2-t_1)\left[\rho(t_1)G^\tn{A}(t_1,t_2)+\int_{t_2}^{t_1}\d t_3\,\int_{t_0}^{t_2}\d  t_3\,G^<(t, t_3)\Sigma^\tn{A}(t_3,  t_4)G^\tn{A}(t_4,t_2)\right]\,.
\end{split}
\end{equation}
To further simplify this equation, they introduced the generalized Kadanoff-Baym ansatz (GKBA), which is equivalent to solving \refeqn{eq:gless_lip} in first order,
 \begin{equation}
\label{eq:gless_gkba}
\begin{split}
G^{(1),<}(t_1,t_2) &= \int_{t_2}^{t_1}\d t_3\,\int_{t_0}^{t_2}\d  t_4\, G^\tn{R}(t_1, t_3)\Sigma^{(0),<}(t_3,  t_4)G^\tn{A}(t_4,t_2)\\ 
&+\Theta(t_1-t_2)\left[-G^\tn{R}(t_1,t_2)\rho^{(0)}(t_2)+\int_{t_2}^{t_1}\d t_3\,\int_{t_0}^{t_2}\d  t_4\,G^\tn{R}(t_1, t_3)\Sigma^\tn{R}(t_3,t_4) G^{(0),<}(t_4,t_2)\right]\,,\\
&+\Theta(t_2-t_1)\left[\rho^{(0)}(t_1)G^\tn{A}(t_1,t_2)+\int_{t_2}^{t_1}\d t_3\,\int_{t_0}^{t_2}\d  t_3\,G^{(0),<}(t, t_3)\Sigma^\tn{A}(t_3,  t_4)G^\tn{A}(t_4,t_2)\right]\,,
\end{split}
\end{equation}
with the result
\begin{align}
\label{eq:GKBAapr}
G^{(0),<}(t_1,t_2)&=\Theta(t_1-t_2)\left[-G^\tn{R}(t_1,t_2)\rho^{(0)}(t_2)\right]+\Theta(t_2-t_1)\left[\rho^{(0)}(t_1)G^\tn{A}(t_1,t_2)\right]\,,\\
\rho^{(0)}(t_1)&=-\i\,G^{(0),<}(t_1,t_1)\,.
\end{align}
One notices that \refeqn{eq:gless_gkba} is only formally closed in terms of the single-particle density matrix $\rho$, since the propagators $G^\tn{R/A}$ still obey a two-time equation of a similar degree of complexity as the original KBE. To circumvent this we introduce a further approximation, replacing the full propagators by the HF propagators $G^\tn{R/A}_\tn{HF}$, which are defined as
\begin{equation}
\label{eq:propagators}
G^\tn{R/A}_\tn{HF}(t_1,t_2)=\mp\i\theta[\pm(t_1-t_2)]\exp\left(-\i\int_{t_2}^{t_1} \d  t_3\,h( t_3)\right),
\end{equation}
Correspondingly, this approximation will be called HF-GKBA.

Then the numerical solution of \refeqn{eq:gless_gkba} can be obtained, using a finite time step $\Delta$, in the following way \cite{stan09},
\begin{equation}
\label{eq:scheme2}
\begin{split}
G^<(t+\Delta,t+\Delta)&=U(t)G^<(t,t)U^\dagger(t)\\
&-\i \Delta\,U(t)I^<(t,t)U^\dagger(t)-\i \Delta\,U(t) \left[I^<(t,t)\right]^\dagger U^\dagger(t)\,,
\end{split}
\end{equation} 
where
\begin{equation}
\begin{split}
\label{eq:U}
X(t+\Delta,t)&=\exp\left(-\i\int_t^{t+\Delta}\dint{}{\bar{t}}\,h(\bar{t})\right) = e^{-\i h(t)\Delta}=:U(t)\;,
\end{split}
\end{equation}
for a small time step $\Delta \ll 1$, and the HF-Hamiltonian $h$ is assumed not to change between $t$ and $t+\Delta$.
The collision integral $I(t,t)$ is given by
\begin{equation}
\label{eq:ITT}
\begin{split}
I(t,t)=\int_{t_0}^t\dint{}{\bar{t}}\,\left\{\Sigma^{>,0}(t,\bar{t})G^{<,0}(\bar{t},t)-\Sigma^{<,0}(t,\bar{t})G^{>,0}(\bar{t},t)\right\}\,,
\end{split}
\end{equation}
where in all the two-time quantities under the integral the GKBA reconstruction according to \refeqn{eq:GKBAapr} is used. 

When comparing \refeqn{eq:ITT} with the right-hand side of \refeqn{eq:neom}, one immediately recognizes that it is of exactly the same structure, with the identifications
\begin{equation}
\begin{split}
\Sigma^{\gtrless ,0}(t,\bar{t}) &=\Sigma_1^\gtrless (t,\bar{t})\,,\\
G^\tn{R}(t,\bar{t}) &= -\Theta(t-{\bar t})U^+(t,\bar{t}) \,,\\
G^{\gtrless ,0}(t,\bar{t}) &=g^{\gtrless} (t,\bar{t})\,.
\end{split}
\end{equation}
Moreover, the GKBA, Eq.~(\ref{eq:GKBAapr}) appeared naturally in our density operator theory, cf. Eq.~(\ref{eq:gkba-do}).
While this equivalence holds for general propagators $U^+$ defined by Eq.~(\ref{eq:DysonU}), neglect of the renormalization (selfenergy term)---which is equivalent to decoupling the BBGKY-hierarchy by using $c_{123}=0$---directly 
leads to HF-propagators and to the HF-GKBA of nonequilibrium Green's functions theory.

\section{Numerical example: 4-site Hubbard model}
\label{sec:results}
To illustrate the findings obtained in sections \ref{sec:DOT}~and~\ref{sec:NEGF}, in this section we show some results for a 4-site Hubbard nano-cluster obtained from the Green's function method with the GKBA applied and compare to density operator results by  \textit{Akbari~et~al.} \cite{akbari12}.
The Hubbard model \cite{hubbard63,lieb03} is a commonly used simplified description of a narrow-band solid-state system, where the motion of the electrons in the solid is mapped onto a hopping process between adjacent atomic sites with just one orbital for each spin projection. The Coulomb interaction between the electrons is assumed to be shielded so that it is mainly restricted to on-site interaction. Despite of these simplifications, this description is able to cover much of the rich behavior of these systems, for example the phase transition between a conductor and Mott insulator through the interplay of hopping and on-site interaction in two and three dimensions. Also in 1D, the system is---depending on the interaction strength---strongly coupled and, therefore, the theoretical description needs to treat correlations between the electrons.

The Hamiltonian of a one-dimensional Hubbard cluster comprised of $N$ sites at electronic half-filling is given by
\begin{equation}
\begin{split}
\label{eq:HubbardH}
\hat H = -t \sum_{ij}^N\sum_\alpha h_{ij}\,{\hat c^\dagger}_{i\alpha}{\hat c}_{j\alpha}+U\sum_{i}^N{\hat c^\dagger}_{i\uparrow}{\hat c}_{i\uparrow}{\hat c^\dagger}_{i\downarrow}{\hat c}_{i\downarrow}\,, 
\end{split}
\end{equation}
where ${\hat c^{(\dagger)}}_{i\alpha}$ denotes the annihilation (creation) operator in the single-particle orbital on site $i$ with spin $\alpha$ and $h_{ij}=\,\delta_{\avg{i,\,j}}$ is the hopping matrix between nearest neighbor sites with the convention, $\delta_{\avg{i,\,j}} = 1$, if  $(i,j)$ are nearest neighbors, and $\delta_{\avg{i,\,j}}=0$, otherwise. The first term originates from the single-particle energies in the periodic lattice structure and incorporates the hopping amplitude $-t$. The second term describes the on-site interaction of the electrons, which is given in terms of the interaction strength $U$.

\begin{figure}[t]
\includegraphics[width=13cm]{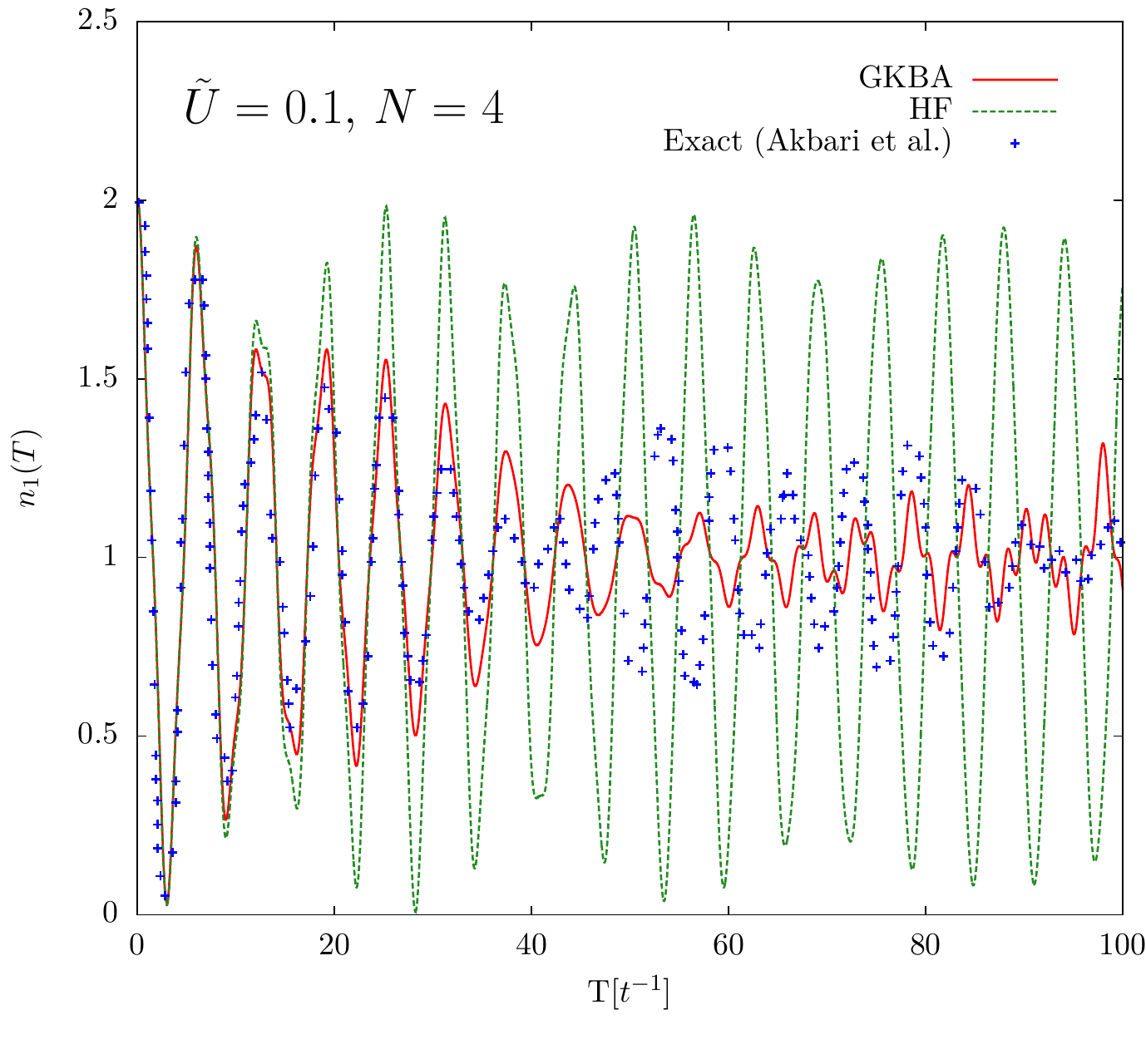}
\caption{Density response on the first site $n_1$ from strong nonequilibrium initial state for the $4$-site Hubbard chain with a coupling strength $\tilde{U}=0.1$. The GKBA and the HF results are depicted by the solid red lines and green dashed lines, respectively. In blue the exact results from \textit{Akbari~et~al.} \cite{akbari12} is shown.}
\label{fig:SNeq1}
\end{figure}

\begin{figure}[t]
\includegraphics[width=13cm]{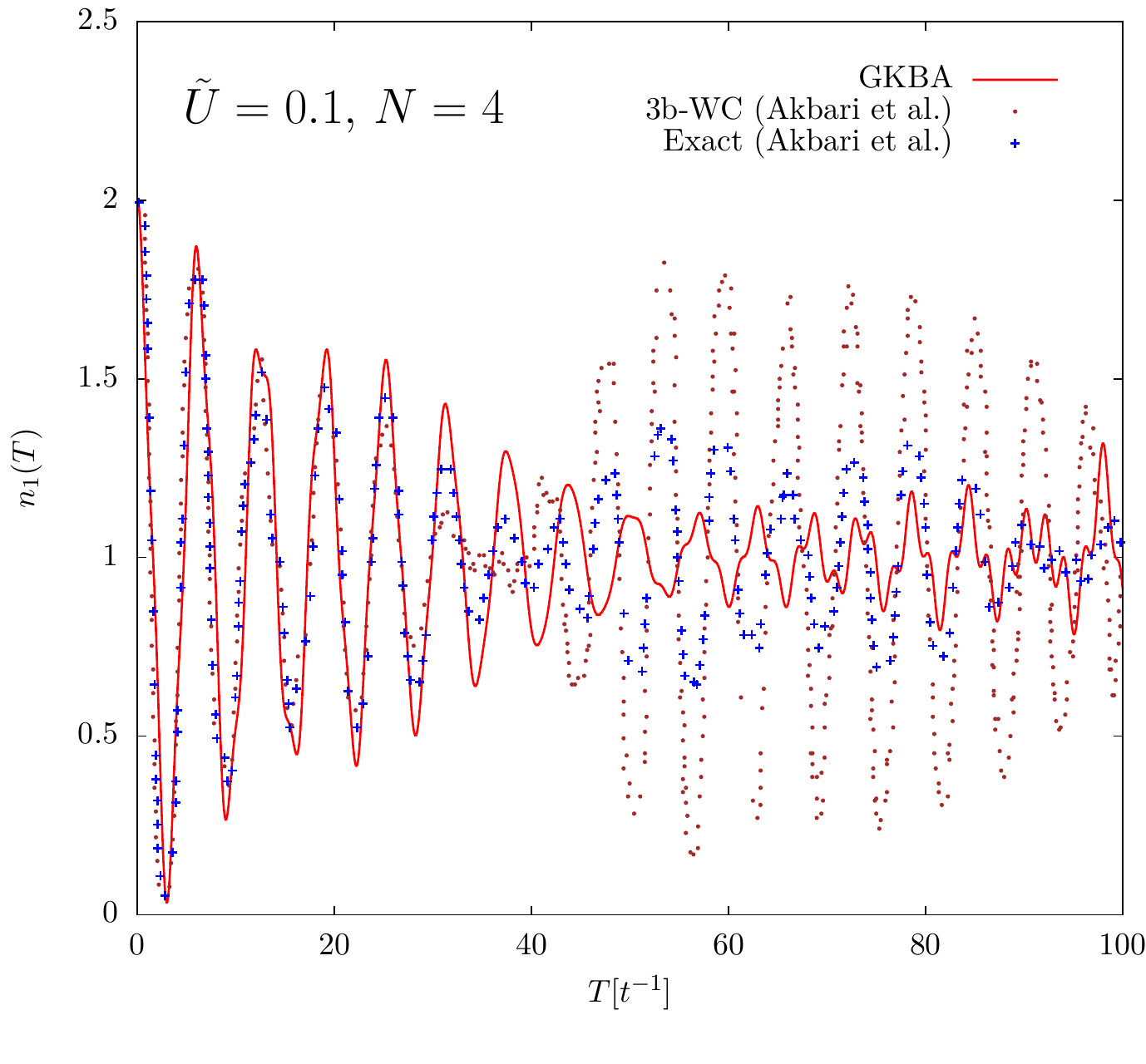}
\caption{Density response on the first site $n_1$ from strong nonequilibrium initial state for the $4$-site Hubbard chain with a coupling strength $\tilde{U}=0.1$. The GKBA results are depicted by the solid red lines. The blue pluses and brown dots represent the exact results and the ones obtained from the WC approximation to the BBGKY hierarchy by \textit{Akbari~et~al.} \cite{akbari12}.}
\label{fig:SNeq2}
\end{figure}

In the following, we consider a $(N=4)$-site cluster with weak interaction, $U=0.1$, at zero temperature with periodic boundary conditions. For this system we are interested in the study of strong non-equilibrium situations. To this end, we prepare the system in an initial state, where all the 4 electrons are forced to the left-most two sites and afterwards examine the free evolution of the system.
For this setup, Green's function solutions within the HF-GKBA, as well as HF results are easily achieved. In \reffigs{fig:SNeq1}{fig:SNeq2}, the time-evolution of the density $n_1(T)$ on the first site---summed over the two orbitals for different spin orientations---measured in terms of the inverse hopping amplitude $t^{-1}$ are shown. In \reffig{fig:SNeq1} we present HF-GKBA results together with the exact solutions and time-dependent HF simulations. In \reffig{fig:SNeq2} 
the HF-GKBA data are compared to exact results as well as to density operator results \cite{akbari12} within the Wang-Cassing (WC) decoupling of the BBGKY-hierarchy \cite{cassing92}. 

The results show that the HF-GKBA performs very well for the propagation of a strong nonequilibrium initial state, especially in comparison with time-dependent Hartree-Fock, that does not show a decrease of the oscillation amplitude at all, present in the exact solution. Comparing the amplitudes with the exact ones, the GKBA overestimates them, though. In the part from $T=40$ on, the GKBA cannot describe the oscillations sufficiently well, but always keeps a good phase agreement with the exact solution. Comparing with the WC approximation in \reffig{fig:SNeq2}, it is obvious that the WC solution has a very good agreement of the oscillation frequency up to $T=80$, but the height of the peaks is much overestimated for time $T>40$, both compared to the exact ones as well as to those computed with the GKBA.       
\section{Discussion}
In this paper we have shown that, within the second order Born approximation, we can establish a one-to-one correspondence between nonequilibrium Green's functions within the GKBA and reduced density operator theory, yielding the same formulas for the equation of motion of the single-particle density matrix. 
On the example of the free evolution of a four-electron quantum dot from a strong nonequilibrium initial state we have shown the overall satisfactory agreement of the HF-GKBA to the exact dynamics of the system far away from equilibrium, for weak interaction strength. This is encouraging since the HF-GKBA allows for long propation times that are impossible to achieve with full two-time calculations and it retains most of the attractive properties of NEGF, such as time reversibility, total energy conservation and memory effects, e.g. \cite{bonitz_pla96,bonitz98}.
In the future, it will be very interesting to examine the potential of the GKBA for higher order selfenergy schemes such as $GW$- or $T$-matrix (ladder) approximation. For the T-matrix approximation for homogeneous macroscopic systems, a similar correspondence has been found previously \cite{kremp97,bonitz98}. It will be interesting to extend this correspondence to inhomogeneous finite systems in order to achieve long propagations also for strongly coupled systems and to verify whether the GKBA is here able, as well, to remove the unphysical damping observed in two-time calculations \cite{puigvonfriesen09}.
\ack
This work is supported in part by the Deutsche Forschungsgemeinschaft via project BO1366/9 and by a grant for CPU time at the HLRN.
\section*{References}
\bibliographystyle{iopart-num}
\providecommand{\newblock}{}

\end{document}